\documentclass[letterpaper,twocolumn,english,aps,multicolumn]{revtex4}
\usepackage[T1]{fontenc}
\usepackage[latin1]{inputenc}

\usepackage{graphics}

\makeatletter

\providecommand{\LyX}{L\kern-.1667em\lower.25em\hbox{Y}\kern-.125emX\@}
\newcommand{\noun}[1]{\textsc{#1}}
\newcommand{\boldsymbol}[1]{\mbox{\boldmath $#1$}}

\makeatother
\begin{document}

\title{Dipolar interactions and anisotropic magnetoresistance in metallic
granular systems}

\author{J. Viana Lopes\thanks{
Author to whom correspondence should be addressed 
}, J. M. B. Lopes dos Santos and Yu. G. Pogorelov}

\address{Centro de Física do Porto and Departamento de Física, Faculdade de
Ciências, Universidade do Porto,\\
 Rua do Campo Alegre, 687, 4169-007 Porto, Portugal}

\begin{abstract}
We revisit the theory of magnetoresistance for a system of nanoscopic
magnetic granules in metallic matrix. Using a simple model for the
spin dependent perturbation potential of the granules, we solve Boltzmann
equation for the spin dependent components of the non equilibrium
electronic distribution function. For typical values of the geometric
parameters in granular systems, we find a peculiar structure of the
distribution function of conduction electrons, which is at variance
with the two-current model of conduction in inhomogeneous systems.
Our treatment explicitly includes the effects of dipolar correlations
yielding a magnetoresistance ratio which contains, in addition to
the term proportional to the square of uniform magnetization \( \left\langle {\boldsymbol \mu }\right\rangle  \),
a weak anisotropic contribution depending on the angle between electric
and magnetic fields, and arising from the anisotropic character of
dipolar interactions.
\end{abstract}
\maketitle

\section{Introduction}

\label{Int}Granular magnetic systems displaying giant magnetoresistance
(GMR) effect \cite{Berk,Xiao}, were put forward as an alternative
to the previously known magnetic multilayers \cite{Baibich,Grunberg},
due to their practical advantages of easier fabrication and higher
stability. On the other hand, the relevant physics in granular systems
presents greater challenges than in the case of multilayered systems,
for which classical \cite{Camley}, semiclassical \cite{Valet} or
quantum \cite{Vedyaev} solutions are available. Of course, in all
these cases the negative GMR has the same origin in that the conduction
electrons with two spin polarizations flow easier at increasing alignment
of localized magnetic moments. But a difficulty results from the fact
that, while the magnetic state of a multilayered system is described
by only few classical variables (the orientations of magnetization
in each magnetic layer), that of a granular system involves a statistical
ensemble of nanoscopic granule magnetic moments which generally cannot
be considered independent. Also the magnetotransport in 3D granular
systems cannot be reduced to any of the paradigmatic cases in the
layered systems, current-in-plane (CIP) or current-perpendicular-to-plane
(CPP) \cite{Rubin}. Nevertheless, currently accepted treatments of
magnetotransport in inhomogeneous materials (both granular and multilayered)
share a common concept: the two-current model, in which the spin up
and spin down electron subsystems carry current independently \cite{Camb,zhang,Wang-Xiao,Allia}.
This model reasonably reproduces the main features of GMR in granular
materials. The most common one is the proportionality of the magnetoresistance
ratio \( \Delta \rho /\rho  \) to the square of uniform magnetization,
\( \Delta \rho /\rho \propto m^{2} \), as first observed by Xiao
\emph{et al} \cite{Xiao} and theoretically explained by assuming
the granules to have a single size and to be uncorrelated \cite{zhang}.
The observed proportionality of \( \Delta \rho  \) to the inverse
of granule diameter \( d \) \cite{Wang-Xiao,Wendong Wang} is accounted
in these models by assuming that the spin dependent scattering in
mainly at the surface of the granules. Experiments also show deviations
from \( m^{2} \) behavior \cite{Xiao,Allia,Ferrari,Hickey} which
have been attributed to granule size dispersion \cite{zhang,Ferrari,yuri},
and/or correlations between the granular moments \cite{Allia,Altbir,yuri,El Hilo,Sousa},
most pronounced when the applied magnetic field is not too strong
compared with the intergranule (dipolar) interactions. 

However, despite appearances, there are some issues open for discussion
in the current accounts of magnetotransport in granular materials.

The treatments referred to above usually consider the scattering from
individual impurities inside the granules to be \emph{incoherent},
despite the fact that electronic mean-free path can be much larger
than granule diameter or even inter granule distance \cite{Wang-Xiao}.
In the present paper we extend a model put forward by Pogorelov \emph{et
al} \cite{yuri} in which the scattering was considered \emph{coherent}
from the whole granule volume and the geometrical factors arising
from this coherence where properly taken into account, albeit within
the context of two-current model. We now perform a full treatment
of the kinetic Boltzmann equation and find that, within this framework,
the two-current model does not hold. In the presence of dominantly
forward scattering, spin flip processes become a crucial determining
factor of the electronic distribution function. In this model the
\( 1/d \) dependence of \( \Delta \rho  \) is obtained without assuming
specific surface scattering. 

The effect of magnetic correlations on transport is usually treated
in the assumption that the resistance is proportional to \( \left\langle {\boldsymbol \mu }_{1}\cdot {\boldsymbol \mu }_{2}\right\rangle  \)
\cite{Allia,Altbir,El Hilo} (where \( {\boldsymbol \mu }_{1} \)
and \( {\boldsymbol \mu }_{2} \) are the magnetic moments of granules),
a phenomenological result that can be traced back to Gittleman \emph{et
al} \cite{Gittleman}\emph{.} The coherent scattering model of Ref.
\cite{yuri} put this result on a firmer footing. Treating in this
model the inverse relaxation time as a spin dependent tensor, determined
by a squared Born scattering amplitude of conduction electrons by
magnetic granules, one readily gets a correlation term \( \left\langle {\boldsymbol \mu }_{1}\cdot {\boldsymbol \mu }_{2}\right\rangle  \)
due to coherent spin-dependent scatterings by the moments \( {\boldsymbol \mu }_{1,2} \)
of two different granules. Also this theory determines the proper
weighting, with distance, of the contribution of this correlation
to transport. Other known approaches to magnetic correlations have
been either phenomenological \cite{Allia}, or numerical \cite{Altbir, El Hilo, Kech}. 

In the present work we are able to calculate the magnetic correlation
function analytically for temperatures well above the characteristic
dipolar energy, believed to be the dominant interaction in these systems
\cite{altbir-2}. These correlations are included in the full treatment
of the Boltzmann kinetic equation. Because dipolar interactions are
not isotropic, we predict a weak dependence of \( \Delta \rho  \)
on the angle \( \theta _{h} \) between the electrical and magnetic
fields. This should not be confused with the anisotropic magnetoresistance
(AMR) seen in systems with a ferromagnetic percolating cluster or
arising from lattice distortions due to film substrate stresses \cite{Mendes,Si}. 

In Sec. \ref{Mod} we discuss the physical parameters for the considered
system and the limits for validity of the related model. In Sec. \ref{Bol}
the spin-dependent Boltzmann equation for this model is formulated
and in Sec. \ref{Sol} the explicit solution is obtained for the distribution
function, emphasizing the importance of spin flip scattering in a
situation where forward scattering dominates. The expressions for
resistivity \( \rho  \) and magnetoresistance ratio \( \Delta \rho /\rho  \)
with various observable dependencies (including a weak dependence
of \( \Delta \rho /\rho  \) on the angle \( \theta _{h} \)) are
presented in section \ref{Rate}. Finally, a general discussion and
some comparisons to previous theories and experimental results are
presented in section \ref{conclusao}.

Partial preliminary results have been already reported in Refs. \cite{jviana,jlopes},
giving a general formulation of the model and the expression for GMR
in a particular geometry (parallel electric and magnetic fields).
Here we give a more detailed presentation of the solution of Boltzmann
equation, generalize it to any angle between electric and magnetic
fields and present a full discussion of the implications of our results.

\section{Definition of physical system and model}

\label{Mod}We consider a metallic system of identical magnetic spherical
granules of diameter \( d \), randomly embedded with volume fraction
\( f \) into non-magnetic metallic matrix. This is a reasonable approximation
to real granular alloys, like Fe\emph{Cu}\emph{\noun{,}} Co\emph{Au}\emph{\noun{,}}
etc., where the magnetic component by transition (T) metal is mostly
aggregated in standard granules (having sizes only slightly dispersed
around the mean value and shapes of polyhedra faceted along low-index
crystalline planes, close to spheres). Conductance in such systems
is mainly realized by \( s \)-like electrons, shared between noble
metal (N) matrix and granules, and they are scattered by magnetic
\( d \)-electrons only present in the granules (it should be noted
that both T- and N-atoms occupy the sites in a common crystalline
lattice). At room temperature, \( d \)-electrons are distributed
in split spin sub-bands and provide almost saturated, uniform magnetization
within each granule. However this magnetization can be randomly oriented
in different granules. Since Fermi \( s \)-electron has in general
different velocities in N- and T-metals and its spin is subjected
to some contact interaction with the polarized spins of \( d \)-electrons
(whole bands), the scattering operator can be modeled by the form
\cite{yuri,jlopes,jviana}:

\begin{equation}
\label{pot}
W_{\sigma \sigma '}(\mathbf{r})=\sum _{j}\chi (\mathbf{r}-\mathbf{R}_{\mathbf{j}})\left( U\delta _{\sigma \sigma \prime }+I{\boldsymbol \tau }_{\sigma \sigma \prime }\cdot {\boldsymbol \mu }_{j}\right) .
\end{equation}
 Here \( U \) and \( I \) are the parameters of potential and spin-dependent
scattering, \( {\boldsymbol \tau }_{\sigma \sigma '} \) is the \( s \)-electron
spin operator. The relevant variable for \( j \)th granule, the unit
vector \( {\boldsymbol \mu }_{j} \) along its magnetic moment, is
considered classical and invariable at electron scattering events,
since the net granule moment \( \mu _{0} \) typically amounts to
\( \sim 10^{4}\mu _{{\textrm{B}}} \) and its coupling to the environment
should be stronger than an energy transfer at electron spin flip.
Thus the model, Eq. \ref{pot}, implies transitions between different
spin channels due to spin precession in the field of classical magnetization
within a granule, rather than due to less probable spin flips by individual
atomic scatterers. The function \( \chi ({\textbf {r}}-{\textbf {R}}_{j}) \)
in the simplest approximation is \( 1 \) when \( \left| {\textbf {r}}-{\textbf {R}}_{j}\right| <r_{0} \),
(\( r_{0} \) being the radius of a granule) and zero otherwise. It
is the Fourier transform of this function (see below) that effectively
accounts for a distinguished role of interface scattering, while it
was to be specially introduced into the incoherent scattering schemes.
The scattering potential, Eq. \ref{pot}, is also easily generalized
to different sized granules. A similar model, which also considered
coherent scattering by the granules, was formerly proposed by Kim
\emph{et al} \cite{Kim}. However, these authors assumed a dipolar
(instead of exchange) coupling between granule magnetization and charge
carriers spin and did not include magnetic correlations between granules.

In what follows, some important relations will be used between the
characteristic length scales for this problem: the Fermi wavelength
\( \lambda _{{\mathrm{F}}}=2\pi /k_{{\mathrm{F}}} \), the mean granule
diameter \( d \), the mean intergranule distance \( D=\left( \pi /6f\right) ^{1/3}d \),
and the mean free path \( \ell  \) for conduction electrons. Namely,
we consider them to obey the following sequence of inequalities

\begin{equation}
\label{ineq}
\ell \gg D>d\gg \lambda _{{\mathrm{F}}}
\end{equation}
 which is rather realistic for experimental systems (see for instance
\cite{Wang-Xiao}). 

A particular physical consequence of the relation \( d\gg \lambda _{{\mathrm{F}}} \)
is that scattering is dominated by small angles as follows from standard
diffraction arguments. We shall see that in this situation the spin
flip scattering has increased importance in determining the structure
of the stationary electronic distribution function.

\section{Boltzmann kinetic equation}

\subsection{Spin-dependent distribution functions }

\label{Bol}We use the description of non-equilibrium electronic state
of a granular system, related to the scattering potential, Eq. \ref{pot},
and subjected to external electric and magnetic fields, in terms of
spin-dependent distribution functions \( f_{\mathbf{k}\sigma } \)
obeying the kinetic Boltzmann equation (BE) \begin{equation}
\label{Beq}
\frac{\partial f_{\mathbf{k}\sigma }}{\partial t}-{\mathbf{v}_{\mathbf{k}}\cdot \nabla _{\mathbf{r}}}f_{\mathbf{k}\sigma }-\frac{e}{\hbar }\left( {\mathbf{E}}+\frac{1}{c}{\mathbf{v}_{\mathbf{k}}}\times {\mathbf{H}}\right) \cdot {\nabla _{\mathbf{k}}}f_{\mathbf{k}\sigma }+
\end{equation}
 \[
+\sum _{\mathbf{k}',\sigma '}\left( f_{\mathbf{k}'\sigma '}-f_{\mathbf{k}\sigma }\right) W_{\mathbf{k}\sigma ,\mathbf{k}'\sigma '}=0\]
 where \( \mathbf{v}_{\mathbf{k}} \) is the conduction electron velocity
and \( W_{\mathbf{k}\sigma ,\mathbf{k}'\sigma '} \) its transition
probability from \( \mathbf{k}'\sigma ' \) to \( \mathbf{k}\sigma  \)
state per unit time. In absence of fields, a trivial steady state
solution holds \( f_{\mathbf{k}\sigma }\equiv f_{\mathbf{k}}^{0}=\{\exp [\beta (\varepsilon _{\mathbf{k}}-\varepsilon _{{\mathrm{F}}})]\}^{-1} \),
describing the spin degenerate Fermi sphere. The electric field distorts
the Fermi surface (FS) by shifting the Fermi sphere and the scattering
redistributes electrons, a stationary distribution resulting from
a balance between these two mechanisms. The current density by two
spin channels is given by \begin{equation}
\label{cur}
{\mathbf{j}}=-e\sum _{\mathbf{k},\sigma }{\mathbf{v}_{\mathbf{k}}}f_{\mathbf{k}\sigma },
\end{equation}
 and in \emph{absence of spin flip scattering} the up and down spin
FS are independent. In this case, the \( \sigma  \) spin FS distortion
is proportional to \( \tau _{\sigma } \), the corresponding relaxation
time, and the conductivity is therefore proportional to \( \tau _{\uparrow }+\tau _{\downarrow } \),
a result used in \cite{zhang,Wang-Xiao,yuri} and in all calculations
in the context of two-current model. Evidently, scattering between
identical spin states is ineffective in relaxing the distortion of
the FS if the angle of scattering is small. The characteristic transport
factor \( 1-\cos \theta  \) in our case can be estimated as\[
1-\cos \theta \approx \frac{\theta ^{2}}{2}\sim \left( \lambda _{{\mathrm{F}}}/d\right) ^{2},\]
 since the diffraction angle \( \theta \approx \lambda _{{\mathrm{F}}}/d\ll 1 \)
.

However, if the spin flip scattering is present, it contributes very
effectively, without the \( 1-\cos \theta  \) factor, to the relaxation
of the \emph{difference} between the up and down spin FS distortions.
So, when scattering in mostly in the forward direction, the spin flip
scattering forces the up and down spin FS distortions to be almost
identical. One then finds that the transport time is given by \[
\frac{1}{\tau _{\textrm{tr}}}=\frac{1}{2}\left( \frac{1}{\tau _{\uparrow }}+\frac{1}{\tau _{\downarrow }}\right) \]
 \emph{i.e.} \emph{the rates,} and not the times, must be averaged.
We also include the effect of correlations between granule magnetic
moments in a way that the scattering kernel of BE involves the connected
correlation functions \begin{equation}
\label{concor}
C_{\alpha }({\mathbf{q}})=\left\langle \sum _{j(\neq i)}C_{ij}^{\alpha }{\textrm{e}}^{-i\mathbf{q}\cdot \mathbf{R}_{ji}}\right\rangle _{{\mathrm{R}}}
\end{equation}
 where \( C_{ij}^{\alpha }=\langle {\boldsymbol \mu }_{i}^{\alpha }\cdot {\boldsymbol \mu }_{j}^{\alpha }\rangle -\langle {\boldsymbol \mu }_{i}^{\alpha }\rangle \cdot \langle {\boldsymbol \mu }_{j}^{\alpha }\rangle , \)
\( \langle \cdots \rangle  \) denotes the thermal average and \( \langle \cdots \rangle _{{\mathrm{R}}} \)
the average over granule positions. Correlations induced by the dipolar
interaction are not isotropic and depend on the angle between \( \mathbf{q} \)
and the external magnetic field. As will be shown below, this results
in a certain dependence of the magnetoresistance on the angle between
current and magnetic field.

\subsection{Solution of kinetic equation}

\label{Sol}In presence of fields, we define a usual expansion to
linearize BE \[
f_{\mathbf{k}\sigma }\equiv f_{\mathbf{k}}^{0}-\frac{\partial f_{\mathbf{k}}^{0}}{\partial \varepsilon _{\mathbf{k}}}\phi _{\mathbf{k}\sigma }\approx f^{0}(\varepsilon _{\mathbf{k}}-\phi _{\mathbf{k}\sigma }),\]
 that is \( \phi _{\mathbf{k}\sigma } \) can be interpreted as the
FS deformation. Then Eq. \ref{cur} is expressed as\begin{equation}
\label{cur1}
{\mathbf{j}}=-2e\sum _{\mathbf{k}}{\mathbf{v}_{\mathbf{k}}}\left( -\frac{\partial f_{\mathbf{k}}^{0}}{\partial \varepsilon _{\mathbf{k}}}\right) \phi _{\mathbf{k}}.
\end{equation}
 where \( \phi _{\mathbf{k}}\equiv \frac{1}{2}\sum _{\sigma }\phi _{\mathbf{k}\sigma } \).
Now BE, Eq. \ref{Beq}, for spatially uniform steady state and with
neglect of orbital effects by magnetic field, can be reduced to an
integral equation for \( \phi _{\mathbf{k}\sigma } \)\begin{equation}
\label{int}
e{\mathbf{v}_{\mathbf{k}}}\cdot {\mathbf{E}}=\sum _{\sigma '}\int d\Omega _{{\mathbf{k}'}}\omega _{\sigma \sigma '}(\mathbf{k},\mathbf{k}')\phi _{\mathbf{k}'\sigma '}.
\end{equation}
 The angular integration in Eq. \ref{int} is over FS and the kernel
\( \omega _{\sigma \sigma '}(\mathbf{k},\mathbf{k}') \) is\begin{eqnarray}
\omega _{\sigma \sigma '}(\mathbf{k},\mathbf{k}') & = & \sum _{\sigma ''}\int \frac{d\Omega _{\mathbf{k}''}}{4\pi }\rho _{\mathrm{F}}\Gamma _{\sigma \sigma ''}(\mathbf{k},\mathbf{k}'')\delta (\Omega _{\mathbf{k}}-\Omega _{\mathbf{k}'})\delta _{\sigma \sigma '}\nonumber \\
 & - & \frac{1}{4\pi }\rho _{\mathrm{F}}\Gamma _{\sigma \sigma '}(\mathbf{k},\mathbf{k}').\label{om10} 
\end{eqnarray}
 where \( \rho _{{\textrm{F}}} \) is the Fermi density of states.
The transition probability density \( \Gamma _{\sigma \sigma '}(\mathbf{k},\mathbf{k}') \)
can be written like in Refs. \cite{yuri,jlopes,jviana} using the
Fermi's golden rule\begin{equation}
\label{Fermi}
\frac{1}{V}\Gamma _{\sigma \sigma '}({\mathbf{k},\mathbf{k}'})=\frac{2\pi }{\hbar }\left| \langle {\mathbf{k}},\sigma |\hat{W}|{\mathbf{k}'},\sigma '\rangle \right| ^{2}
\end{equation}
 where \( V \) is the sample volume.

For the common case of point-like scatterers, the validity of Born
approximation only requires that the relevant energy scales for perturbation,
\( U \) and \( I \), be small compared to the Fermi energy \( \varepsilon _{{\mathrm{F}}} \)
\cite{AGD}. But in our case, the finite size of scatterers also needs
some consideration. In a na\"ive view, the strength of perturbation
could be represented by the scattering rates (see below)\begin{equation}
\label{gammaU-I}
\gamma _{U}^{2}=\frac{2\pi }{\hbar }V_{0}U^{2}\rho _{{\mathrm{F}}},\qquad \gamma _{I}^{2}=\frac{2\pi }{\hbar }V_{0}I^{2}\rho _{{\mathrm{F}}},
\end{equation}
 where \( V_{0}=\pi d^{3}/6 \) is the granule volume. For a relevant
choice of parameters: \( V_{0}\sim10  \) nm\( ^{3} \), \( U\sim I\sim 0.3 \)
eV, \( \rho _{{\mathrm{F}}}\sim 10 \) eV\( ^{-1} \)nm\( ^{-3} \),
\( \varepsilon _{{\mathrm{F}}}\sim 5 \) eV, one has a large ratio
\( \hbar \gamma _{I}^{2}/\varepsilon _{{\mathrm{F}}}>1 \). However,
this estimate ignores the fact that the scattering is mostly in forward
direction. Taking this into account, the lifetime of a momentum state
is estimated as : \[
\hbar /\tau \sim f(\gamma _{U}^{2}+\gamma _{I}^{2})/(k_{{\mathrm{F}}}d)^{2}\approx f(U^{2}+I^{2})k_{{\mathrm{F}}}d/\varepsilon _{{\mathrm{F}}}.\]
 Hence the condition of weak scattering \( \hbar /\tau \ll \varepsilon _{{\mathrm{F}}} \)
remains valid even for values of \( U \) and \( I \) as large as
\( \sim 1 \) eV.

Then BE, Eq. \ref{int}, can be rewritten as an operator equation
in the space of functions defined on FS,\begin{equation}
\label{oper}
\sum _{\sigma '}\hat{\omega }_{\sigma \sigma '}|\phi _{\sigma '}\rangle =|\phi _{{\mathrm{E}}}\rangle 
\end{equation}
 where \( |\phi _{{\mathrm{E}}}\rangle  \) denotes the driving term
and \( |\phi _{\sigma }\rangle  \) the FS distortion for spin \( \sigma  \).

We use the angular momentum basis \( |\ell m\rangle  \) and define
the coefficients\[
\phi _{\ell m\sigma }\equiv \langle \ell m|\phi _{\sigma }\rangle \equiv \phi _{\ell m}+\sigma \eta _{\ell m},\]
 so that \[
\phi _{{\mathbf{k}}\sigma }=\sum _{\ell ,m}\phi _{\ell m\sigma }{\textrm{Y}}_{\ell }^{m}({\hat{\mathbf{k}}}).\]
 Note that since the driving term of BE is \( e{\mathbf{v}_{\mathbf{k}}\cdot \mathbf{E}}\propto ev_{\mathbf{k}}E{\textrm{Y}}^{0}_{1}(\hat{\mathbf{k}}) \),
only the \( \ell =1,m=0 \) component of \( |\phi _{{\mathrm{E}}}\rangle  \)
is non zero. As was already mentioned, correlations induced by dipolar
interactions lead to a dependence of the scattering kernel on the
angle between the momentum transfer vector \( \mathbf{q}\equiv \mathbf{k}'-\mathbf{k} \)
and the external magnetic field, and it is convenient to separate
out its (small) anisotropic part from the main isotropic one:\begin{eqnarray}
\Gamma _{\sigma \sigma '}(\mathbf{k},\mathbf{k}') & = & \Gamma _{\sigma \sigma '}^{(i)}(q)+\Gamma ^{(a)}_{\sigma \sigma '}(\mathbf{q}),\label{i-a} \\
\hat{\omega }_{\sigma \sigma '}(\mathbf{k},\mathbf{k}') & = & \hat{\omega }^{(i)}_{\sigma \sigma '}(q)+\hat{\omega }^{(a)}_{\sigma \sigma '}({\mathbf{q}}).\nonumber 
\end{eqnarray}
 Let us first restrict consideration to the isotropic scattering kernel,
\( \omega ^{(i)}_{\sigma \sigma '}(q) \), which depends only on the
angle between \( \mathbf{k} \) and \( \mathbf{k}' \). The resulting
operator in Eq. \ref{oper} is diagonal in the spherical harmonic
basis with \( z \) axis along the only distinguished direction of
electric field \( \mathbf{E} \) (E-basis). Note that in this case
the quantization axis for electron spin operator \( {\widehat{\boldsymbol \tau }} \)
can be chosen arbitrary. Then, presenting the relevant solution as
\( \phi _{10\sigma }^{(i)}=\phi ^{(i)}_{10}+\sigma \eta ^{(i)}_{10} \)
we obtain its components \( \phi ^{(i)}_{10} \) and \( \eta ^{(i)}_{10} \)
from\begin{eqnarray}
 & \omega ^{(i)}_{ss}\phi ^{(i)}_{10}+ & \omega _{sa}^{(i)}\eta ^{(i)}_{10}=\sqrt{4\pi /3}ev_{\mathbf{k}}E,\label{BE10} \\
 & \omega ^{(i)}_{as}\phi ^{(i)}_{10} & +\omega _{aa}^{(i)}\eta ^{(i)}_{10}=0,\nonumber 
\end{eqnarray}
 all other components of the FS distortion being zero. We should emphasize
that, in any case, only the \( \ell =1 \) components of \( \phi _{\mathbf{k}\sigma } \)
contribute to the current. The rates appearing in Eq. \ref{BE10}
are given by\begin{eqnarray}
\omega ^{(i)}_{ss}=\frac{1}{2}\sum _{\sigma \sigma '}\langle 10|\hat{\omega }_{\sigma \sigma '}^{(i)}|10\rangle ,\: \: \: \: \: \:  &  & \label{om-s-a} \\
\omega ^{(i)}_{sa}=\frac{1}{2}\sum _{\sigma \sigma '}\sigma '\langle 10|\hat{\omega }_{\sigma \sigma '}^{(i)}|10\rangle ,\: \:  &  & \nonumber \\
\omega ^{(i)}_{as}=\frac{1}{2}\sum _{\sigma \sigma '}\sigma \langle 10|\hat{\omega }_{\sigma \sigma '}^{(i)}|10\rangle ,\: \: \:  &  & \nonumber \\
\omega ^{(i)}_{aa}=\frac{1}{2}\sum _{\sigma \sigma '}\sigma \sigma '\langle 10|\hat{\omega }_{\sigma \sigma '}^{(i)}|10\rangle , & \nonumber 
\end{eqnarray}
 or, more explicitly,\begin{eqnarray*}
\omega ^{(i)}_{ss}=\frac{\rho _{\mathrm{F}}}{2}\sum _{\sigma ,\sigma '}\int \frac{d\Omega _{{\mathbf{k}'}}}{4\pi } & \Gamma ^{(i)}_{\sigma \sigma '}(|{\mathbf{k}}-{\mathbf{k}'}|) & (1-\cos (\theta _{\mathbf{kk}'})),\\
\omega ^{(i)}_{sa}=\frac{\rho _{\mathrm{F}}}{2}\sum _{\sigma ,\sigma '}\int \frac{d\Omega _{{\mathbf{k}'}}}{4\pi } & \Gamma ^{(i)}_{\sigma \sigma '}(|{\mathbf{k}}-{\mathbf{k}'}|) & (\sigma -\sigma '\cos (\theta _{\mathbf{kk}'})),\\
\omega ^{(i)}_{as}=\frac{\rho _{\mathrm{F}}}{2}\sum _{\sigma ,\sigma '}\int \frac{d\Omega _{{\mathbf{k}'}}}{4\pi } & \Gamma ^{(i)}_{\sigma \sigma '}(|{\mathbf{k}}-{\mathbf{k}'}|) & \sigma (1-\cos (\theta _{\mathbf{kk}'})),\\
\omega ^{(i)}_{aa}=\frac{\rho _{\mathrm{F}}}{2}\sum _{\sigma ,\sigma '}\int \frac{d\Omega _{{\mathbf{k}'}}}{4\pi } & \Gamma ^{(i)}_{\sigma \sigma '}(|{\mathbf{k}}-{\mathbf{k}'}|) & (1-\sigma \sigma '\cos (\theta _{\mathbf{kk}'})).
\end{eqnarray*}
 Solving Eqs. \ref{BE10}, we obtain \( \phi ^{(i)}_{10}=\sqrt{4\pi /3}ev_{\mathbf{k}}E\tau ^{(i)}_{tr} \)
where \begin{equation}
\label{tau-tr-1}
\tau ^{(i)}_{tr}\, ^{-1}=\omega ^{(i)}_{ss}-\frac{\omega ^{(i)}_{sa}\omega ^{(i)}_{as}}{\omega ^{(i)}_{aa}}
\end{equation}
 is the transport relaxation time in isotropic approximation. Introducing
\( \phi _{\mathbf{k}} \) into Eq. \ref{cur1} in the same approximation:
\[
\phi ^{(i)}_{\mathbf{k}}=\phi ^{(i)}_{10}{\textrm{Y}}_{1}^{0}(\hat{\mathbf{k}})=e{\mathbf{v}_{\mathbf{k}}\cdot \mathbf{E}}\tau ^{(i)}_{tr},\]
 we arrive in a standard way at the Drude resistivity \[
\rho =\frac{m_{e}}{n_{e}e^{2}\tau ^{(i)}_{tr}}.\]
 As stated in Sec. \ref{Mod} and explicitly shown below, the scattering
probability \( \Gamma _{\sigma \sigma '}(\mathbf{k},\mathbf{k}') \)
is dominated by small angles \( \theta _{\mathbf{kk}'} \). As a result,
the integrals in Eqs. \ref{om-s-a} that involve the factor \( 1+\cos \theta  \)
are larger than those with \( 1-\cos \theta  \) by a factor of \( (k_{\mathrm{F}}d)^{2} \).
They appear only in \( \omega _{aa} \) (and cancel in \( \omega _{sa} \))
so that\begin{equation}
\label{tau_i-app}
\tau ^{(i)}_{tr}\, ^{-1}\approx \omega _{ss}^{(i)}.
\end{equation}
 Considering the definitions of Eq. \ref{om-s-a}, we can easily conclude
that \begin{equation}
\label{om-ss}
\omega _{ss}^{(i)}=\frac{1}{2}\left( \tau ^{(i)}_{\uparrow \uparrow }\, ^{-1}+\tau ^{(i)}_{\downarrow \downarrow }\, ^{-1}+2\tau ^{(i)}_{\downarrow \uparrow }\, ^{-1}\right) 
\end{equation}
 where the rates \( \tau ^{(i)}_{\sigma \sigma '}\, ^{-1} \) are
defined as \[
\tau ^{(i)}_{\sigma \sigma '}\, ^{-1}=\rho _{\mathrm{F}}\int \frac{d\Omega _{{\mathbf{k}'}}}{4\pi }\Gamma ^{(i)}_{\sigma \sigma '}(|{\mathbf{k}-\mathbf{k}'}|)(1-\cos (\theta _{\mathbf{kk}'})).\]
 Eq. \ref{om-ss} expresses the fact, already discussed in Sec. \ref{Bol},
that up and down spin FS have almost the same deformation and so the
relaxation rate of the mean of the two deformations is just the mean
of corresponding rates. For a typical value \( k_{\mathrm{F}}d\approx 40 \)
the neglected term in Eq. \ref{tau-tr-1} turns out to be about \( 0.5\% \)
of the term retained.

The weak anisotropic term of the scattering kernel, \( \hat{\omega }^{(a)}_{\sigma \sigma '} \),
due to the correlation between magnetic moments, can be easily included
into the present treatment in a perturbative way. Then Eq. \ref{oper}
reads\[
\sum _{\sigma '}\left( \hat{\omega }_{\sigma \sigma '}^{(i)}|\phi _{\sigma '}\rangle +\hat{\omega }_{\sigma \sigma '}^{(a)}|\phi _{\sigma '}\rangle \right) =|\phi _{{\mathrm{E}}}\rangle ,\]
 and the solution can be written as \( |\phi _{\sigma }\rangle =|\phi ^{(i)}_{\sigma }\rangle +|\phi ^{(a)}_{\sigma }\rangle  \),
where \( |\phi ^{(a)}_{\sigma }\rangle  \) is a small anisotropic
perturbation. To lowest non zero order in it, we have \begin{equation}
\label{BE-ia}
\sum _{\sigma '}\left( \hat{\omega }_{\sigma \sigma '}^{(i)}|\phi ^{(a)}_{\sigma }\rangle +\hat{\omega }_{\sigma \sigma '}^{(a)}|\phi ^{(i)}_{\sigma '}\rangle \right) =0.
\end{equation}
 Projecting out the \( \ell =1 \) and \( m=0 \) components and using
the facts that \( \hat{\omega }_{\sigma \sigma '}^{(i)} \) is diagonal
and the only non zero component in the unperturbed solution \( |\phi ^{(i)}_{\sigma }\rangle  \)
is that with \( \ell =1,\, m=0 \), we can solve this equations for
the coefficients \( \phi ^{(a)}_{10\sigma }=\langle 10|\phi ^{(a)}_{\sigma }\rangle =\phi ^{(a)}_{10}+\sigma \eta ^{(a)}_{10} \).
A simplification similar to that leading to Eq. \ref{tau_i-app} also
applies here and we obtain \[
\phi _{10}^{(a)}=-\frac{\omega _{ss}^{(a)}}{\omega ^{(i)}_{ss}}\phi ^{(i)}_{10}\]
 where \( \omega ^{(a)}_{ss} \) is defined in a similar way to \( \omega ^{(i)}_{ss} \)
in Eq. \ref{om-s-a}. Hence the transport time \( \tau _{tr} \) is
changed by a factor \( 1-\omega ^{(a)}_{ss}/\omega ^{(i)}_{ss} \)
compared to \( \tau ^{(i)}_{tr} \), Eq. \ref{tau_i-app}, and to
the order of accuracy that we are working in Eq. \ref{BE-ia}, we
may write \begin{equation}
\label{tau-tr-2}
\tau ^{-1}_{tr}=\omega ^{(i)}_{ss}+\omega ^{(a)}_{ss}.
\end{equation}
 One might question at this point, whether it is legitimate to include
this last correction while neglecting the second term in Eq. \ref{tau-tr-1}.
It should be stressed however that the two terms in Eq. \ref{tau-tr-2}
are of the same order with respect to the small parameter \( 1/(k_{\mathrm{F}}d) \).
So our theory is consistently a lowest non zero order theory in this
small parameter. On the more practical side, we will see that for
typical parameter values this correction, arising from spin correlations,
can in fact be more important than the terms neglected in Eq. \ref{tau-tr-1}.

\section{Calculation of transport rates\label{Rate}}

\subsection{Isotropic Kernel}

The principles for the calculation of the scattering rates have already
been spelled out in Ref. \cite{yuri}. The explicit squaring matrix
element in Eq. \ref{Fermi} is\begin{equation}
\label{square}
\left| \langle {\mathbf{k}},\sigma |\hat{W}|{\mathbf{k}'},\sigma '\rangle \right| ^{2}=\frac{V^{2}_{0}}{V^{2}}\psi ^{2}(\frac{qd}{2})\left| \sum _{j}e^{-i{\mathbf{q}}\cdot {\mathbf{R}}_{j}}W^{j}_{\sigma \sigma '}\right| ^{2}
\end{equation}
where the function \( \psi (x)=3(\sin (x)-x\cos (x))/x^{3} \) is
the structure factor of a sphere, the Fourier image of the function
\( \chi (r) \), and \( W^{j}_{\sigma \sigma '}=U\delta _{\sigma \sigma '}+I{\boldsymbol \tau }_{\sigma \sigma '}\cdot {\boldsymbol \mu }_{j} \).
It is due to the presence of factors \( \psi ^{2}(qd/2) \) in the
integrals of Eq. \ref{om-s-a} that only values of \( q\leq d^{-1} \)
contribute significantly to the scattering. Evidently, Eq. \ref{square}
should be averaged with respect to the random positions \( {\textbf {R}}_{i} \)
of the granules and over the thermal distribution of their moments
\( {\boldsymbol \mu }_{i} \). The latter can be done in the basis
with \( z \) axis along the external magnetic field \( \mathbf{H} \)
(H-basis), where only the \( z \) component of a magnetic moment
has non zero average \( \langle \mu _{i}^{z}\rangle \equiv \langle \mu _{z}\rangle  \).

We separate out contribution involving a single granule, and write
averages of products of magnetic moments of diferent granules as products
of averages plus connected correlation functions, Eq. \ref{concor}.
It is then straightforward to obtain the explicit formulas for diagonal
and non-diagonal components of isotropic and anisotropic parts of
the scattering rates, Eq. \ref{i-a}:\begin{eqnarray}
\rho _{\mathrm{F}}\Gamma ^{(i)}_{\sigma \sigma }(q) & = & f\psi ^{2}(\frac{qd}{2})\Big (\gamma _{U}^{2}\left( 1+g(q)\right) +\nonumber \\
 & + & 2\sigma \gamma _{U}\gamma _{I}\langle \mu _{z}\rangle \left( 1+g(q)\right) \label{gamma1} \\
 & + & \gamma _{I}^{2}\left( \langle \mu _{z}^{2}\rangle +g(q)\langle \mu _{z}\rangle ^{2}\right) \Big ),\nonumber \\
\rho _{\mathrm{F}}\Gamma ^{(i)}_{\sigma ,-\sigma }(q) & = & f\psi ^{2}(\frac{qd}{2})\gamma _{I}^{2}\left[ 1-\langle \mu _{z}^{2}\rangle \right] ,\nonumber 
\end{eqnarray}
and\begin{eqnarray}
\rho _{\mathrm{F}}\Gamma ^{(a)}_{\sigma \sigma }({\mathbf{q}})=f\psi ^{2}(\frac{qd}{2})\gamma _{I}^{2}C_{\parallel }(\mathbf{q}),\qquad  &  & \label{gamma2} \\
\rho _{\mathrm{F}}\Gamma ^{(a)}_{\sigma ,-\sigma }({\mathbf{q}})=f\psi ^{2}(\frac{qd}{2})\gamma _{I}^{2}C_{\perp }(\mathbf{q}). &  & \nonumber \label{gamma3} 
\end{eqnarray}
 Here \( g(q) \) is the pair correlation function, already calculated
in Refs. \cite{yuri,jviana,jlopes}, using the excluded volume approximation\begin{eqnarray*}
g(q) & \equiv  & \frac{1}{N}\langle \sum _{i\neq j}e^{-i\mathbf{q}\cdot (\mathbf{R}_{j}-\mathbf{R}_{i})}\rangle _{\{R\}}\approx \\
 & \approx  & \frac{f}{V_{0}}\int _{r>d}d\mathbf{r}\textrm{ e}^{-i\mathbf{q}\cdot \mathbf{r}}=-8f\psi (qd)
\end{eqnarray*}
 (a delta function at \( q=0 \) is neglected because it gives a zero
contribution to the integrals over \( q \)). The correlation functions
\( C_{\parallel }({\mathbf{q}})\equiv C_{z}({\mathbf{q}}) \) and
\( C_{\perp }({\mathbf{q}})\equiv C_{x}({\mathbf{q}})+C_{y}({\mathbf{q}}) \)
and the relevant average \( \langle \mu _{z}\rangle  \) are calculated
in the same approximation in Appendix B.

The angular integrations appearing in Eq. \ref{om-s-a} can now be
transformed to integrals over the momentum transfer \( q \). They
can be classified in terms of inverse powers of the large parameter
\( k_{\mathrm{F}}d \) (see Appendix A). The terms, involving the
factor \( 1+\cos \theta  \), are of the order \( (\gamma _{U}^{2},\gamma _{I}^{2})f/(k_{\mathrm{F}}d)^{2} \)
whereas the factors involving the factor \( 1-\cos \theta \propto q^{2} \)
are smaller by a factor \( 1/(k_{\mathrm{F}}d)^{2} \) \emph{i.e.,}
of the order \( (\gamma _{U}^{2},\gamma _{I}^{2})f/(k_{\mathrm{F}}d)^{4} \).

In the absence of correlations the result for the transport rate is
quite simple (see Eq. \ref{tau-tr-1})

\begin{eqnarray}
\tau _{tr}^{-1} & = & \frac{4f}{(k_{\mathrm{F}}d)^{4}}\left( \alpha \gamma ^{2}-\beta f\gamma _{I}^{2}\langle \mu _{z}\rangle ^{2}\right) \label{tau-tr-3} 
\end{eqnarray}
 where \( \gamma ^{2}=\gamma _{U}^{2}+\gamma _{I}^{2} \) and the
constants \( \alpha =(9/2)\ln (k_{\mathrm{F}}d) \), and \( \beta \approx 4.172 \)
are defined in Appendix A. Other (non-magnetic) scattering mechanisms
will give rise to an additive background contribution to the resistivity
\( \rho _{b} \) so that \[
\rho =\rho _{b}+\frac{m_{e}}{n_{e}e^{2}}\frac{4f}{(k_{\mathrm{F}}d)^{4}}\left( \alpha \gamma ^{2}-\beta f\gamma _{I}^{2}\langle \mu _{z}\rangle ^{2}\right) \]
 This result is identical to the first two terms of Eq. (18) in Ref.
\cite{yuri} and it gives a magnetoresistance proportional to the
square of magnetization\begin{equation}
\label{gmr}
-\Delta \rho =\frac{m_{e}}{n_{e}e^{2}}\frac{4\beta f^{2}\gamma ^{2}_{I}}{(k_{\mathrm{F}}d)^{4}}\langle \mu _{z}\rangle ^{2}.
\end{equation}
 Such proportionality of the magnetoresistance to \( \langle \mu _{z}\rangle ^{2} \),
at this level of approximation is also obtained in all calculations
in the two-current model \cite{zhang,Wang-Xiao}. Differences between
these approaches will be discussed in more detail in Sec. \ref{conclusao}.

\subsection{Contribution of Correlations}

\label{Cor}A calculation of the effect of correlations on transport
requires explicit expressions for the correlation functions. In this
article we consider only the high temperature limit of dipolar interactions,
\( k_{\mathrm{B}}T>\mu _{0}^{2}/D^{3} \) when the high temperature
expansion is meaningful. The calculation gives (see Appendix B)\begin{eqnarray}
C({\mathbf{q}}) & = & \frac{8\pi f\mu ^{2}_{0}}{3V_{0}k_{\mathrm{B}}T}{\mathcal{L}}_{2}(\frac{\mu _{0}h}{k_{\mathrm{B}}T})\psi (qd)P_{2}(\cos \theta _{{\mathbf{q},\mathbf{h}}})\label{corr} 
\end{eqnarray}
 where \( {\mathcal{L}}_{2}(s)=[{\mathcal{L}}(s)/s]^{2}-[{\mathcal{L}}'(s)]^{2} \)
and \( {\mathcal{L}}(s)=\coth (s)-1/s \) is the Langevin function.
The \( {\mathcal{L}}_{2} \) factor is related to the common field
effect on the magnetization. A more subtle field effect on the transport
follows from the factor \( P_{2}(x)\equiv (3x^{2}-1)/2 \), depending
on the angle \( \theta _{{\mathbf{q},\mathbf{h}}} \) between the
scattering vector and the external magnetic field. This dependence
results from the calculation of the correlator \( C({\mathbf{q}}) \)
using the H-basis (that where \( z \)-axis is along \( \mathbf{H} \)).
It eventually introduces a dependence of the transport time on the
angle \( \theta _{h} \) between the current and the magnetic field
(between E- and H-bases). Returning again to the E-basis (with \( \mathbf{H} \)
lying in the \( xz \) plane) for integration in Eq. \ref{om10} and
using the results of Eq. \ref{gamma1}, one readily arrives at the
following expression for the relevant rate \( \omega ^{(a)}_{ss} \):
\begin{eqnarray}
\omega _{ss}^{(a)}=\frac{2}{3}\frac{f^{2}\gamma _{I}^{2}\mu _{0}^{2}}{V_{0}k_{\mathrm{B}}T}{\mathcal{L}}_{2}(\frac{\mu _{0}h}{k_{\mathrm{B}}T})\times \qquad \qquad \qquad \qquad  &  & \label{omass} \\
\times \int d\Omega _{\mathbf{k}}\int d\Omega _{\mathbf{k}'}\psi ^{2}(\frac{qd}{2})\psi (qd)P_{2}(\cos \theta _{{\mathbf{q},\mathbf{h}}})\times  &  & \nonumber \\
\times \left[ |Y^{0}_{1}(\theta _{\mathbf{k}})|^{2}-Y_{1}^{0}(\theta _{\mathbf{k}'})Y^{0}_{1}(\theta _{\mathbf{k}})\right] . &  & \nonumber 
\end{eqnarray}
 The arguments of \( \psi  \) functions contain \( q=2k_{\mathrm{F}}\sin (\theta _{\mathbf{kk}'}/2) \)
were \( \theta _{\mathbf{kk}'} \) is the angle between \( \mathbf{k} \)
and \( \mathbf{k}' \). The integrals over \( \Omega _{\mathbf{k}} \)
and \( \Omega _{\mathbf{k}'} \) in Eq. \ref{omass} are calculated
using the addition theorem for spherical harmonics to give

\begin{eqnarray*}
\omega _{ss}^{(a)} & = & -\frac{14\pi }{15}\frac{\beta f^{2}\gamma ^{2}_{I}\mu _{0}^{2}}{V_{0}k_{\mathrm{B}}T(k_{\mathrm{F}}d)^{4}}{\mathcal{L}}_{2}(\frac{\mu _{0}h}{k_{\mathrm{B}}T})P_{2}(\cos \theta _{h}).
\end{eqnarray*}
 So, finally we obtain for the transport rate\begin{eqnarray}
\rho  & = & \rho _{b}+\frac{m_{e}}{n_{e}e^{2}}\frac{4f}{(k_{\mathrm{F}}d)^{4}}\left\{ \Bigg .\alpha \gamma ^{2}-\right. \label{rho_novo} \\
 &  & \left. -\beta f\gamma _{I}^{2}\left[ \langle \mu _{z}\rangle ^{2}+\frac{7\pi \mu _{0}^{2}}{30V_{0}k_{\mathrm{B}}T}{\mathcal{L}}_{2}(\frac{\mu _{0}h}{k_{\mathrm{B}}T})P_{2}(\cos \theta _{h})\right] \Bigg .\right\} \qquad \nonumber 
\end{eqnarray}
 where again a background contribution, \( \rho _{b} \), from non-magnetic
scattering mechanisms (other impurities, phonons) is included. The
magnetoresistance is then modified from Eq. \ref{gmr} to\begin{equation}
\label{eq-gmr}
-\Delta \rho =\frac{m_{e}}{n_{e}e^{2}}\frac{4\beta f^{2}\gamma ^{2}_{I}}{(k_{\mathrm{F}}d)^{4}}\left( \langle \mu _{z}\rangle ^{2}+\frac{7\pi }{30}\frac{T_{0}}{T}{\mathcal{L}}_{2}(\frac{\mu _{0}h}{k_{\mathrm{B}}T})P_{2}(\cos \theta _{h})\right) 
\end{equation}
 where \( T_{0} \), the characteristic temperature of dipolar interactions,
is \begin{equation}
\label{T0}
T_{0}=\frac{\mu _{0}^{2}}{V_{0}k_{\mathrm{B}}}.
\end{equation}
The last term in Eq. \ref{eq-gmr} describes the deviation from \( \sim \langle \mu _{z}\rangle ^{2} \)
behavior of GMR due to dipolar magnetic correlations.

\section{Discussion and Conclusions}

\label{conclusao}When the coherence between individual scattering
events inside a granule is disregarded \cite{Berk,zhang,Wang-Xiao}
the relaxation rate of spin \( \sigma  \) electrons \( \Delta _{\sigma }\equiv \tau ^{-1}_{\sigma }=\Delta _{0}+\sigma \Delta _{1} \)
is proportional to the volume fraction of granules, \( f \), and
\( \Delta _{1} \) is proportional to the mean normalized magnetic
moment \( \left\langle \mu _{z}\right\rangle  \). Furthermore, if
magnetic scattering at the granule surface dominates, \( \Delta _{1} \)is
proportional to the surface to volume ratio of the granules \emph{i.e.}
inversely proportional to \( d \), the diameter. In the two-current
model the spin up and down electron subsystems carry current independently
and the conductivity is given by \[
\sigma =\frac{n_{e}e^{2}}{m_{e}}(\tau _{\uparrow }+\tau _{\downarrow })\]
or \[
\rho (h)=\frac{m_{e}}{n_{e}e^{2}}\frac{\Delta ^{2}_{0}-\Delta ^{2}_{1}}{\Delta _{0}}.\]
As a result the magnetoresistance is \[
-\Delta \rho =\frac{m_{e}}{n_{e}e^{2}}\frac{\Delta ^{2}_{1}}{\Delta _{0}}.\]
From this a proportionality follows of \( \Delta \rho  \) to \( \left\langle \mu _{z}\right\rangle ^{2} \),
\( 1/d \) and \( f \), provided the resistance in zero field is
also dominated by surface scattering at the granules (\( \Delta _{0} \)
proportional to \( f \) and \( 1/d \) ).

In our model the entire granule scatters coherently. We believe this
is reasonable, in view of the fact that the electronic mean free path
can be larger than the granule diameter or even the inter-granular
distance \cite{Wang-Xiao}. This corresponds to the homogeneous limit
of Camblong \emph{et al} \cite{Camb}. Given the large size of the
granules, compared to the Fermi wavelength, scattering is predominant
in the forward direction, concentrated in a cone of angular size of
order \( (\lambda _{\mathrm{F}}/d)^{2}\sim 1/(k_{\mathrm{F}}d)^{2} \).
As we emphasized, spin flip scattering then enforces that the up and
down spin FS distortions stay in step, and the resistivity is given
by \[
\rho =\frac{m}{n_{e}e^{2}}(\Delta _{\uparrow }+\Delta _{\downarrow })\]
The interference between scatterings in different granules gives a
contribution to \( \Delta _{\sigma } \) proportional to \( \left\langle {\boldsymbol \mu }_{1}\cdot {\boldsymbol \mu }_{2}\right\rangle  \)
which, in the absence of correlations yields a \( \left\langle \mu _{z}\right\rangle ^{2} \)
term proportional to \( f^{2} \). Since in our calculation surface
scattering is not specifically distinguished, one might expect the
scattering cross section to be proportional to the volume of the granule
\( \sim d^{3} \). However, as we mentioned above, it is decreased
by a factor of \( 1/(k_{\mathrm{F}}d)^{2} \) by a standard diffraction
argument. The contribution to the resistivity carries an extra factor
\( 1/(k_{\mathrm{F}}d)^{2} \) arising from the transport factor \( 1-\cos \theta  \).
As a result the resistivity is proportional to \( 1/d \).

Given that our model corresponds to the homogeneous limit one could
question whether it is capable of predicting the actual values of
magnetoresistance ratio observed in experiments. 

By defining a resistivity scale \[
\Delta \rho _{m}=\frac{m_{e}}{n_{e}e^{2}}\frac{4\beta f^{2}\gamma ^{2}_{I}}{(k_{\mathrm{F}}d)^{4}}\]
we may rewrite Eq. \ref{eq-gmr} in the form\begin{equation}
\label{delta_rho_rm}
-\Delta \rho =\Delta \rho _{m}\left( \langle \mu _{z}\rangle ^{2}+\frac{7\pi T_{0}}{30T}{\mathcal{L}}_{2}(\frac{\mu _{0}h}{k_{\mathrm{B}}T})P_{2}(\cos \theta _{h})\right) .
\end{equation}
The magnetic field dependent factor (within brackets) is zero for
zero field and unity for saturating field, where correlations no longer
contribute, so \( \Delta \rho _{m} \) is the maximum value of \( |\Delta \rho | \).
The zero field resistance can be written as (see eq. \ref{rho_novo})
\begin{equation}
\label{rho-0-rho-m}
\rho (0)=\rho _{b}+\frac{\alpha }{\beta f}\left( 1+\frac{\gamma ^{2}_{U}}{\gamma ^{2}_{I}}\right) \Delta \rho _{m}
\end{equation}
 This gives an upper bound for the magnetoresistance ratio\begin{equation}
\label{maximo}
\frac{\Delta \rho _{m}}{\rho (0)}\leq \frac{\beta f}{\alpha }
\end{equation}
The ratio \( \beta /\alpha \approx 0.94/\ln k_{\mathrm{F}}d \) varies
between \( 0.31\sim 0.23 \) for \( k_{\mathrm{F}}d \) in the range
of \( 20\sim 60 \). This would imply \( |\Delta \rho |/\rho \leq 0.3f \),
that is considerably smaller than what is usually observed below room
temperature. The relatively small value of \( \beta /\alpha  \) is,
however, a consequence of assuming a sharp granule interface in the
function \( \chi ({\textbf {r}}-{\textbf {R}}_{j}) \), which leads
to a logarithmic factor in \( \alpha  \). To check this, we have
also calculated with smoother density profiles and, for values of
the interface thickness as small as 10\% of the diameter \cite{Rabedeau},
a rather realistic \( \beta /\alpha  \) \( \sim 0.8 \) is obtained.
For comparison, Wang and Xiao found maximum magnetoresistance ratios
of the order of 5\% at 300 K and 15 \% at 77 K in a sample of Fe\( _{20} \)
Ag\( _{80} \) with \( d=29 \) \AA~\cite{Wang-Xiao}. These authors
also found \( \rho (0) \) to be inversely proportional to \( d \),
which means that the second term in Eq. \ref{rho-0-rho-m} should
dominate over \( \rho _{b} \). 

The effect of magnetic correlations on transport is contained in the
second term of Eq. \ref{delta_rho_rm}. Our treatment applies to dipolar
interactions in the high temperature limit \( \mu ^{2}_{0}/D^{3}\ll k_{\mathrm{B}}T \),
where \( D \) is a typical minimum distance between granules. Since
\( D^{3}=V_{0}/f, \) this condition is equivalent to \( T\gg fT_{0}. \)
Using, as an example, the values of the sample mentioned above we
estimate \( T_{0}\approx 107 \) K and \( fT_{0}\approx 21 \) K.
This shows that the high temperature approximation used to calculate
the correlator, Eq. \ref{corr}, is actually valid down to reasonably
low temperatures.

\begin{figure}
{\centering \resizebox*{9cm}{!}{\rotatebox{270}{\includegraphics{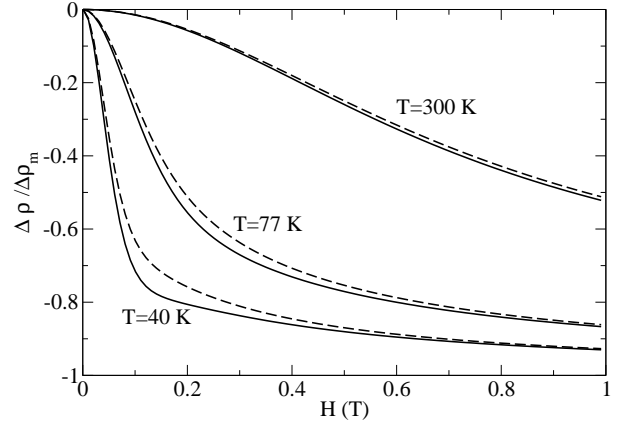}}} \par}

\caption{\label{fig1}Field dependent factor in the magnetoresistance as a
function of magnetic field at various temperatures. Dashed lines -
without dipolar correlation correction; full lines - with inclusion
of dipolar correlation correction at parallel electric and magnetic
fields (\protect\( T_{0}=107\, \textrm{K}\protect \), \protect\( \mu _{0}=1492\, \mu _{\mathrm{B}}\protect \)
and \protect\( f=0.2\protect \) \cite{Wang-Xiao}). }
\end{figure}

To illustrate the effect of the correlation term, we compare in Fig.
\ref{fig1}
the factor \( \langle \mu _{z}\rangle ^{2}+(7\pi T_{0}/30T){\cal L}_{2}(\mu _{0}h/k_{\mathrm{B}}T) \)
for \( \theta _{\mathbf{h}}=0 \) (strongest correlation effect) with
\( \langle \mu _{z}\rangle ^{2} \) alone, as functions of magnetic
field for different temperatures (we used the parameters of the sample
mentioned above, \( T_{0}=107\, \textrm{K} \) and \( \mu _{0}=1492\, \mu _{\mathrm{B}} \)).
As expected, the correlations give a larger correction (as much as
10\%) at lower temperatures. A plot of magnetoresistance ratio \( \Delta \rho /\rho  \)
vs \( \langle \mu _{z}\rangle ^{2} \) should be a straight line if
the correlations are neglected. In Fig. 
\begin{figure}
{\centering \resizebox*{9cm}{!}{\rotatebox{270}{\includegraphics{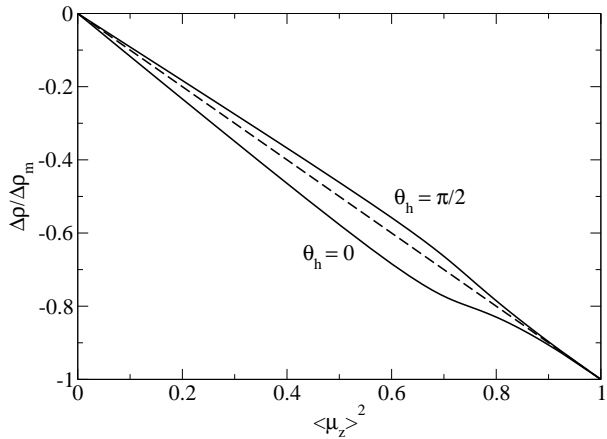}}} \par}

\caption{\label{fig2}Field dependent factor of magnetoresistance ratio, as
a function of \protect\( \langle \mu _{z}\rangle ^{2}\protect \),
for \protect\( T=40\protect \) K and the same sample parameters as
in Fig. \ref{fig1}. The effect of dipolar correlations is seen in
the curvature, which has opposite signs for parallel (\protect\( \theta _{h}=0\protect \)
) and perpendicular (\protect\( \theta _{h}=\pi /2\protect \)) electric
and magnetic fields. Notice that for the angle \protect\( \theta _{h}=\arccos 1/\sqrt{3}\approx 55^{\circ }\protect \)
the dipolar correlation correction turns zero and the straight line
is recovered. }
\end{figure}
 \ref{fig2} we illustrate this effect by representing \( \Delta \rho /\Delta \rho _{m} \)
vs \( \langle \mu _{z}\rangle ^{2} \) (see Eq. \ref{delta_rho_rm})
for parallel (\( \theta _{h}=0 \)) and perpendicular (\( \theta _{h}=\pi /2 \))
electric and magnetic fields. Note that the deviation from linear
behaviour depends on temperature, granule parameters \( \mu _{0} \)
and \( d \), through the temperature scale \( T_{0} \) (Eq. \ref{T0}),
but \emph{not} on the concentration \( f \). Our result is remarkably
similar to the one found by Kechrakos and Trohidou in a numerical
calculation for mono-dispersive samples (see fig. 14 of their paper
\cite{Kech}). This effect, although small, has the interesting feature
of depending on the angle between the electric and magnetic fields
(see Fig.\ref{fig2}) and having opposite signs in the case of parallel
or perpendicular fields. The dependence on the relative orientation
of electric and magnetic fields is due to the anisotropic nature of
the dipolar interactions. The correlation between two granules is
ferromagnetic for the moment components along the direction joining
the granules but antiferromagnetic for the perpendicular components.
The correlation correction vanishes and linearity of \( \Delta \rho /\rho  \)
vs \( \langle \mu _{z}\rangle ^{2} \) is restored for the particular
angle \( \theta _{h}=\arccos 1/\sqrt{3}\approx 55^{\circ } \). These
features, together with a specific temperature dependence, should
facilitate experimental detection of the dipolar correlation effect
in GMR of granular metallic systems. It should be mentioned that granule
size dispersion also leads to a non-linear \( \Delta \rho  \) vs
\( \langle \mu _{z}\rangle ^{2} \) \cite{zhang, yuri, Kech} but
always convex and \( \theta _{h} \)-independent. Both these kinds
of non-linearity can coexist in real granular systems. At last we
note that the correlation term by Eq. \ref{delta_rho_rm} is zero
at zero magnetic field. However, at temperatures lower than \( fT_{0} \),
dipolar correlations cannot be treated within high temperature expansion,
and it may happen then, that the zero field correlation effect remains
finite. This in fact could rise the maximum magnetoresistance above
the limit of Eq. \ref{maximo}, \( \Delta \rho _{m} \). The anisotropy
we predict may already have been noticed in CuAg films by Stearns
and Cheng \cite{Stearns}. It should not be confused with usual AMR,
due to skew magnetic scattering, observed in more concentrated systems
\cite{Mendes}. 

To conclude, we would like to stress once more that these results
follow from a peculiar structure of the electron distribution function,
related to the fact that magnetic scattering is mostly in the forward
direction. The relaxation rate of the difference between the up and
down spin FS is the fastest one, leading to identical up and down
spin FS distortions. The resulting model differs in some ways from
the two-current model of magnetotransport, but seems to account equally
well for the basic features of experiments. Some of the ideas presented
here may also apply in other systems whenever large structures, but
still smaller than electronic mean free path, dominate magnetic scattering.

\section{Acknowledgments}

The authors would like to thank A.V. Vedyaev, B.L. Altshuler, A. H.
Castro Neto, V.L. Belinicher, E. S. Lage and S.N. Dorogovtsev for
very illuminating discussions. This work is supported by Fundação
da Ciência e Tecnologia of Portugal, through project PRAXIS XXI 2/2.1/FIS/302/94.
One of the authors (J. Viana Lopes) is supported by FCT grant \emph{SFRH/BD}/1261/2000.

\section{Appendix A}

The angular integrations, over the direction of \( \mathbf{k}' \),
in Eqs. \ref{om-s-a} can be transformed into integrations over the
momentum transfer \( q=2k_{\mathrm{F}}\sin \theta _{\mathbf{kk}'}/2 \).
The term \( \cos \theta _{\mathbf{kk}'}=1-2(q/2k_{\mathrm{F}})^{2} \)
and the solid angle integration element, \( d\Omega _{\mathbf{k}'}=8\pi qdq/(2k_{\mathrm{F}})^{2} \).
Using of Eqs. \ref{gamma1} in Eqs. \ref{om-s-a} involves the following
integrals:\begin{eqnarray*}
I\left( Q\right) =\int _{0}^{Q}x^{3}\psi ^{2}(x)dx=\quad \qquad \qquad \qquad \qquad \qquad \qquad  &  & \\
=\frac{9}{2}\left[ \ln 2Q+\gamma _{{\mathrm{E}}}-1-{\textrm{Ci}}\left( 2Q\right) +\right. \qquad \qquad  &  & \\
\left. +\frac{\sin 2Q}{Q}-\frac{1-\cos 2Q}{2Q^{2}}\right] ,\qquad  &  & \\
J\left( Q\right) =\int _{0}^{Q}x^{3}\psi ^{2}(x)\psi (2x)dx=\qquad \qquad \qquad \qquad \qquad  &  & \\
=\frac{27}{20}\left[ 2\ln 2-1+2{\textrm{Ci}}\left( 2Q\right) -2{\textrm{Ci}}\left( 4Q\right) -\right. \qquad \qquad  &  & \\
-\frac{\sin 2Q\left( 1-\cos 2Q\right) }{Q}\left( 1+\frac{3}{4Q^{2}}+\frac{1}{4Q^{4}}\right) +\qquad \qquad  &  & \\
\left. +\frac{5+4\cos 2Q-\cos 4Q}{8Q^{2}}+\frac{\cos 2Q-\cos 4Q}{2Q^{4}}\right] , &  & 
\end{eqnarray*}
 where \( \gamma _{{\mathrm{E}}}\approx 0.5772 \) is Euler's gamma.
The transition rates, Eq. \ref{tau-tr-2}, are simply expressed through
these integrals:\begin{eqnarray*}
\tau _{{\textrm{tr}}}^{-1} & = & \frac{4f}{\left( k_{\mathrm{F}}d\right) ^{4}}\Bigg (\gamma ^{2}I\left( k_{\mathrm{F}}d\right) -J(k_{\mathrm{F}}d)8f\gamma _{\mathrm{I}}^{2}\times \\
 &  & \times \Big (\langle \mu _{z}\rangle ^{2}+\frac{7\pi }{30}\frac{T}{T0}{\cal L}_{2}(\frac{\mu _{0}h}{k_{\mathrm{B}}T})P_{2}(\cos \theta _{\textrm{h}})\Big )\Bigg )
\end{eqnarray*}
 and in the limit of \( k_{\mathrm{F}}d\gg 1 \) (taking into account
the asymptotics of integral cosine \( {\textrm{Ci}}(x)\rightarrow \sin x/x \))
we have the result of Eq. \ref{tau-tr-3} with \( \alpha =(9/2)\ln (k_{\mathrm{F}}d) \)
and \( \beta =(54/5)(2\ln 2-1)\approx 4.172 \).

\section{Appendix B }

\label{Fun_cor}The Hamiltonian for magnetic interactions of the granules
is\[
H=-\mu _{0}h\sum _{i}\mu _{i}^{z}+\mu _{0}^{2}\sum _{<ij>\alpha \, \beta }U^{\alpha \beta }_{ij}\mu ^{\alpha }_{i}\mu _{j}^{\beta }\]
 where \( h \) is the external magnetic field applied to the system
(with \( z \) axis rotated compared to the geometry in Sec. \ref{Cor})
and \( U^{\alpha \beta }_{ij}=(\delta _{\alpha \beta }-3u^{\alpha }_{ij}u^{\beta }_{ij})/r_{ij}^{3} \)
(with the unit vector \( {\mathbf{u}}_{ij}={\mathbf{r}}_{ij}/r_{ij} \))
is the dipolar interaction tensor. We made a standard high temperature
expansion, with respect to the dipolar interaction energy, valid for
temperatures for which the dipolar interaction is a small perturbation
to the non-interacting Hamiltonean.

To the first order, we obtain for the magnetic moment \( \langle \mu _{z}\rangle  \)
\[
\left\langle \mu ^{z}_{i}\right\rangle ={\mathcal{L}}(s)\left( 1-\frac{\mu _{0}^{2}}{k_{\mathrm{B}}T}\sum _{j\neq i}U^{zz}_{ij}{\mathcal{L'}}(s)\right) ,\]
and the moment-moment correlation function \( C_{ij}=\left\langle {\boldsymbol \mu }_{i}\cdot {\boldsymbol \mu }_{j}\right\rangle -\left\langle {\boldsymbol \mu }_{i}\right\rangle \cdot \left\langle {\boldsymbol \mu }_{j}\right\rangle  \):\begin{eqnarray*}
C_{ij} & = & U^{zz}_{ij}\left( \frac{\partial {\mathcal{L}}(s)}{\partial s}\right) ^{2}+(U^{xx}_{ij}+U^{yy}_{ij})\left( \frac{{\mathcal{L}}(s)}{s}\right) ^{2}\\
 & = & \frac{1-3\cos ^{2}\theta _{ij}}{r_{ij}^{3}}{\mathcal{L}}_{2}(s),
\end{eqnarray*}
 with \( s=\mu _{0}h/k_{\mathrm{B}}T \). These expressions must still
be averaged over the positions of the granules. We assume a uniform
distribution with the excluded volume constraint \cite{yuri} and
obtain for \( \langle \mu _{z}\rangle  \): \begin{eqnarray*}
\langle \mu _{z}\rangle ={\mathcal{L}}(s)-\beta \mu _{0}^{2}{\mathcal{L}}(s){\mathcal{L'}}(s)\times \qquad \qquad  &  & \\
\times \frac{f}{V_{0}}\int _{|{\mathbf{r}}_{j}-{\mathbf{r}}_{i}|>2r_{0}}d^{3}r_{j}U_{ij}^{zz}, &  & 
\end{eqnarray*}
 which can be written in terms of the classical demagnetizing factor
\( N_{z} \) of the granular sample as\[
\langle \mu _{z}\rangle ={\mathcal{L}}(s)-f\frac{T_{0}}{T}{\mathcal{L}}(s){\mathcal{L'}}(s)\left( N_{z}-\frac{4\pi }{3}\right) .\]
 Finally, the Fourier transform (restricted by the excluded volume)
\( C({\mathbf{q}})\equiv C_{\parallel }({\mathbf{q}})+C_{\perp }({\mathbf{q}}) \)
of the correlation function is \begin{eqnarray*}
C({\mathbf{q}}) & = & \frac{f}{V_{0}}\int _{r>d}d^{3}{\mathbf{r}}C({\mathbf{r}})e^{i{\mathbf{q}}\cdot {\mathbf{r}}}=\\
 & = & \frac{8\pi }{3}f\frac{T_{0}}{T}{\mathcal{L}}_{2}(s)\psi (2qr_{0})P_{2}(\cos \theta _{{\mathbf{q},\mathbf{h}}}).
\end{eqnarray*}

\end{document}